\documentclass[aps,prb,twocolumn,superscriptaddress,nofootinbib]{revtex4-2}
\usepackage[colorlinks=true,citecolor=blue,linkcolor=blue,urlcolor=blue]{hyperref}
\usepackage{graphicx, nicefrac, textcomp}
\usepackage[normalem]{ulem}

\begin{document}

\title{Soft X-ray Spectroscopy of Low-Valence Nickelates}
\author{Matthias~Hepting}
\email[]{hepting@fkf.mpg.de}
\affiliation{Max-Planck-Institute for Solid State Research, Stuttgart, Germany\\}
\author{Mark~P.~M.~Dean}
\affiliation{Condensed Matter Physics and Materials Science Department, Brookhaven National Laboratory, Upton, NY, USA\\}
\author{Wei-Sheng~Lee}
\affiliation{Stanford Institute for Materials and Energy Sciences, SLAC National Accelerator Laboratory and Stanford University, Menlo Park, CA, USA\\}
\date{\today}

\begin{abstract}

Low-valence nickelates — including infinite-layer (IL) and trilayer (TL) compounds — are longstanding candidates for mimicking the high-temperature superconductivity of cuprates. A recent breakthrough in the field came with the discovery of superconductivity in hole-doped IL nickelates. Yet, the degree of similarity between low-valence nickelates and cuprates is the subject of a profound debate for which soft x-ray spectroscopy experiments at the Ni $L$- and O $K$-edge provided critical input. In this review, we will discuss the essential elements of the electronic structure of low-valance nickelates revealed by x-ray absorption spectroscopy (XAS) and resonant inelastic x-ray scattering (RIXS). Furthermore, we will review magnetic excitations observed in the RIXS spectra of IL and TL nickelates, which exhibit characteristics that are partly reminiscent of those of cuprates.

\end{abstract}

\maketitle

\section{Introduction}

Spectroscopy has proven to be a versatile tool for studying the charge, spin, lattice, and orbital degrees of freedom in quantum materials. Arguably, one of the most fascinating families of quantum materials are cuprate high-temperature superconductors \cite{Keimer2015,Lee2006,Armitage2010}. Critical insights into cuprates have been provided by electron, neutron, and photon spectroscopy techniques — including photoemission and electron spectroscopy, scanning tunneling microscopy, optical and Raman spectroscopy, as well as inelastic x-ray and neutron scattering \cite{Damascelli2003,Fink2001,Fischer2007,Basov2005,Devereaux2007,Fujita2012, Miao2018}. More recently, fresh perspectives on the enigmatic ground states of cuprates opened up especially due to resonant soft x-ray spectroscopy studies. For instance, resonant inelastic x-ray scattering (RIXS) \cite{Ament2011} revealed that damped spin excitations persist even for high hole-doping levels far away from the parent antiferromagnetically (AFM) ordered phase \cite{LeTacon2011, Dean2013}. Furthermore, RIXS allowed to probe the three-dimensional (3D) dispersion of low-energy plasmons \cite{Hepting2018,Lin2019,Nag2020}, which arise due to the characteristic quasi-2D layered crystal structure of cuprates. Moreover, resonant elastic x-ray scattering (REXS) and RIXS studies found that different types of static and dynamic charge orders emerge ubiquitously in cuprates \cite{Ghiringhelli2012, Miao2017, Chaix2017, Miao2019a, Lee2021}. Nevertheless, a comprehensive understanding of the most prominent phases in cuprates — including the pseudogap, strange metal phase, and superconductivity — remains elusive.

One approach to gain a deeper understanding of cuprates involves the targeted design of materials that mimic cuprate-typical properties, such as their layered  quasi-2D  crystal structure, 3$d^9$ electronic configuration, spin $S = 1/2$ magnetic moments with antiferromagnetic (AFM) coupling, strong ligand-oxygen hybridization, and a lifted degeneracy of the active $e_g$ orbitals \cite{Norman2016}. In principle, the discovery of superconductivity in a material that emulates at least a subset of these properties could allow to identify the hallmarks of cuprates that are crucial for invoking their exceptional high-temperature superconductivity. 

In this context, long-standing candidates are Ni-based compounds, as Ni is a direct neighbor of Cu in the periodic table. In early works, it was speculated that doped $RE_2$NiO$_4$ ($RE$ = rare-earth ion), which is the $n = 1$ member of the Ruddlesden-Popper (RP) homologous series \textit{RE}$_{n+1}$Ni$_n$O$_{3n+1}$ \cite{Greenblatt1997}, could become superconducting \cite{Acrivos1994}, as it is isostructural to the cuprate La$_2$CuO$_4$ and possesses similar charge and spin stripe ordered states \cite{Sachan1995}. The formal electronic configuration of La$_2$NiO$_4$, however, is 3$d^8$ (Ni$^{2+}$) with $S = 1$ \cite{Sugai1990,Fabbris2017}, providing a possible rationalization for the observed absence of superconductivity.
Perovskite nickelates $RE$NiO$_3$ are the $n = \infty$ member of the RP series with a formal 3$d^7$ (Ni$^{3+}$) configuration and $S = 1/2$, although x-ray spectroscopic experiments indicated a 3$d^8\underline{L}$ configuration \cite{Bisogni2016}, with $\underline{L}$ denoting an O ligand hole. In their seminal work, Chaloupka and Khaliullin proposed a significant modification of the electronic structure of perovskite nickelates via tensile strain and incorporation into thin epitaxial heterostructures \cite{Chaloupka2008},  breaking of the degeneracy of the $e_g$ orbitals, which in the extreme case could yield a half filled 3$d_{x^2-y^2}$ band and a single-sheet, cuprate-like Fermi surface \cite{Hansmann2009}. 
In fact, spectroscopic studies detected new electronic and magnetic ground states in perovskite nickelate heterostructures, distinct from those of bulk nickelates \cite{Middey2016,Catalano2018,Boris2011,Benckiser2011,Liu2011,Frano2013,Hepting2014,Gibert2015,Disa2017,Hepting2018a,Fursich2019}. Nevertheless, superconductivity has not been found to date \cite{Chakhalian2011,Wu2013,Disa2015}, possibly due to the insufficient splitting of the orbital energy levels in the realized heterostructures \cite{Han2011,Fabbris2016}. More recently, oxygen-reduced variants of the $n = 3$ members of the RP series have attracted intense attention. These \textit{RE}$_{4}$Ni$_{3}$O$_{8}$ trilayer (TL) nickelates are composed of three closely stacked square-planar Ni-O layers with intervening $RE$ ions, which are separated by rocksalt type $RE$-O blocking structures \cite{Zhang2016}. The electronic configuration of TL nickelates is 3$d^{8.67}$ per Ni on average (Ni$^{1.33+}$). It was found by x-ray spectroscopy that metallic Pr$_{4}$Ni$_{3}$O$_{8}$ exhibits low-spin configuration and a significantly lifted orbital degeneracy, suggesting a close analogy to cuprates \cite{Zhang2017N}. Notably, the close parallel between TL nickelates and cuprates was further corroborates by RIXS, revealing the presence of strong AFM exchange coupling $J$ in La$_{4}$Ni$_{3}$O$_{8}$ and Pr$_{4}$Ni$_{3}$O$_{8}$ \cite{Lin2021}. Nonetheless, superconductivity has not been detected in TL nickelates. This could be due to their electronic configuration, which corresponds to 1/3-hole-doping of a Ni$^{1+}$ background and therewith falls into the overdoped regime of cuprates \cite{Zhang2017N,Botana2017,Nica2020}. Experimental efforts to stabilize superconductivity via lowering the Ni$^{1.33+}$ valence by electron-doping are ongoing. 
Along these lines, also the $n = 5$ oxygen-reduced RP variants \textit{RE}$_{6}$Ni$_{5}$O$_{12}$ are promising candidates for superconductivity, as these quintuple layer nickelates exhibit a 3$d^{8.8}$ (Ni$^{1.2+}$) configuration — analogous to optimally doped cuprates — already without additional electron-doping. 

A breakthrough in the field came with the discovery of superconductivity in hole-doped nickelates with the IL crystal structure \cite{Li2019}. In more detail, epitaxial thin films of Sr- or Ca-substituted $RE$NiO$_2$ obtained via topotactic oxygen reduction of the perovskite phase show superconductivity below 9 - 15 K \cite{Li2019,Zeng2020,Li2020a,Lee2020,Osada2020,Li2021G,Osada2021}. The parent compounds of these nickelates formally exhibit the 3$d^{9}$ (Ni$^{1+}$) configuration with $S = 1/2$, which qualifies them as isostructural and isoelectronic to the parent cuprates. Early neutron powder diffraction studies of parent IL nickelates, however, indicated absence of long-range AFM order \cite{Hayward1999,Hayward2003} and electrical transport measurements of films show weakly metallic behavior \cite{Ikeda2016}. This is in stark contrast to  parent cuprates, which are AFM Mott (charge-transfer) insulators \cite{Keimer2015}. Moreover, whereas first theoretical studies proposed that the electronic and magnetic correlations of IL nickelates and cuprates share close similarities \cite{Anisimov1999}, other theoretical works suggested significant distinctions, including a multiband character of nickelates \cite{Lee2004}. Along these lines, insights from experiments can help to resolve the controversy about similarities and differences between IL nickelates and cuprates. In particular, recent x-ray and electron energy-loss spectroscopic studies \cite{Hepting2020,Goodge2021,Lu2021} unveiled a reduced Ni-O hybridization, presence of a weakly interacting $RE$ 5$d$ metallic band, and overdamped spin excitations with a bandwidth as large as 200 meV in IL nickelates.

In the following, we will review recent soft x-ray absorption spectroscopy (XAS) and RIXS studies at the O $K$-edge and Ni $L$-edge of IL and TL nickelates. We will discuss the essential elements of their distinct electronic structure. Furthermore, the spin excitation spectra of IL and TL nickelates observed with RIXS will be reviewed.

\section{Electronic structure}

\begin{figure*}[h!]
\begin{center}
\includegraphics[width=17cm]{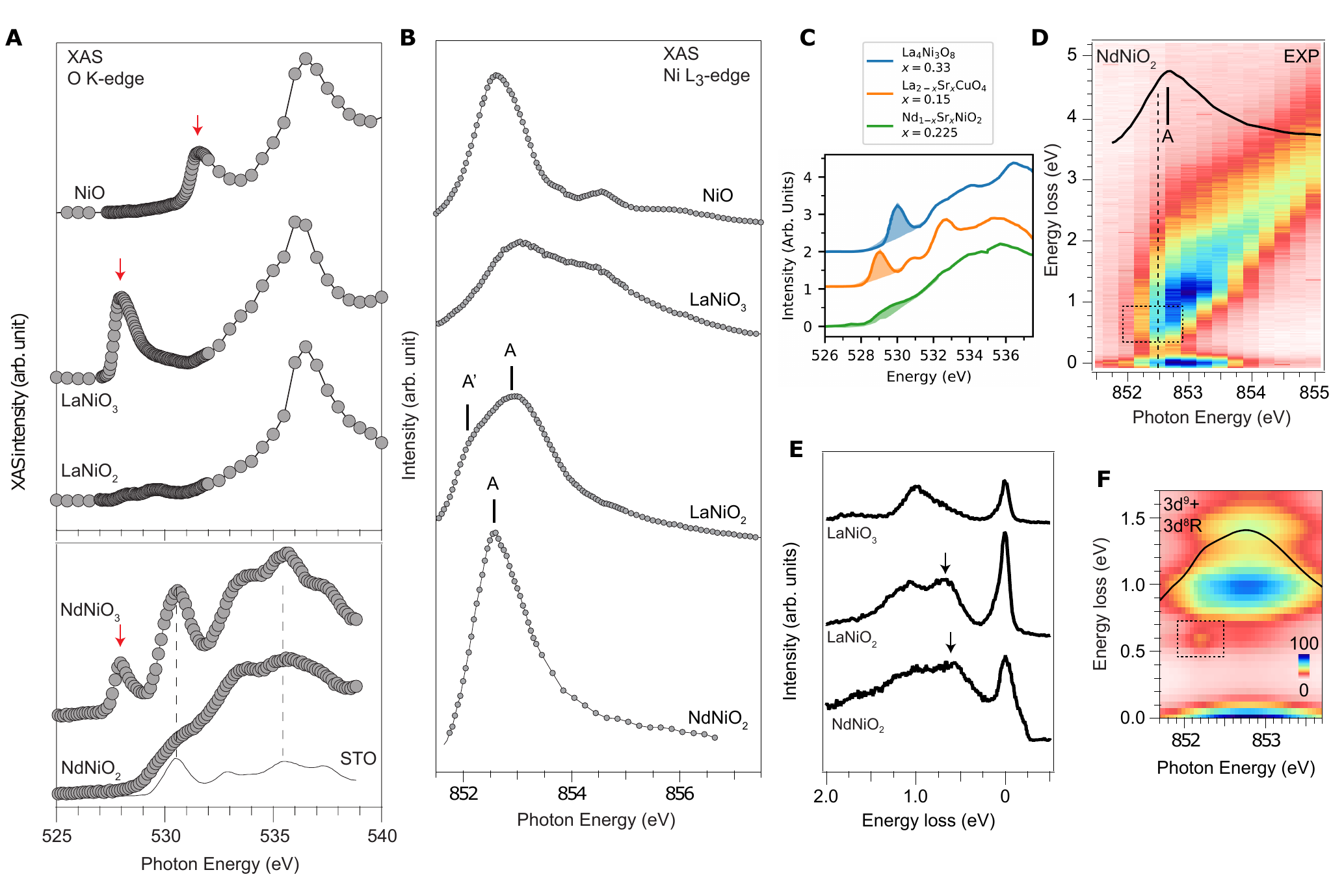}
\end{center}
\caption{Electronic structure of nickelates. \textbf{(A)} Upper panel: O $K$-edge XAS  of NiO, LaNiO$_3$ and LaNiO$_2$. Red arrows mark the pre-edge peaks indicative of Ni–O hybridization. Lower panel: O $K$-edge XAS of NdNiO$_3$ and NdNiO$_2$. Dashed vertical lines indicate features of the SrTiO$_3$
(STO) substrate (solid grey line) in the XAS spectra of NdNiO$_3$ and NdNiO$_2$ due to the film thickness being thinner than that of the La-based films in the upper panel. Spectra are vertically offset for clarity. \textbf{(B)} Ni $L_3$-edge XAS of NiO, LaNiO$_3$, LaNiO$_2$, and NdNiO$_2$. The La $M_4$-line was subtracted from the LaNiO$_3$ and LaNiO$_2$ spectra. \textbf{(C)} Comparison of the O $K$-edge pre-peak intensities (shaded areas) of a TL nickelate (La$_{4}$Ni$_{3}$O$_{8}$), a hole-doped cuprate (La$_{1.85}$Sr$_{0.15}$CuO$_{4}$), and a hole-doped IL nickelate (Nd$_{0.775}$Sr$_{0.225}$NiO$_{2}$). \textbf{(D)} RIXS intensity map of NdNiO$_2$ measured as a function of incident photon energy across the Ni $L_3$-edge. \textbf{(E)} Representative RIXS spectra of LaNiO$_3$, LaNiO$_2$, and NdNiO$_2$. Black arrows highlight the 0.6 eV features of LaNiO$_2$ and NdNiO$_2$. \textbf{(F)} Calculated RIXS map and XAS (solid black line) of LaNiO$_2$ for a 3$d^9$ + 3$d^8R$ ground state, with $R$ denoting a charge-transfer to the La cation. The dashed box highlights the same feature as the box in panel D. Panels adapted from Refs.~\cite{Hepting2020,Lin2021}.}\label{electronic}
\end{figure*}

XAS at the O $K$-edge measures core-hole excitations from O 1$s$ to unoccupied O 2$p$ states and is also a sensitive probe of the covalent mixing between O 2$p$ and transition-metal $d$ states \cite{Abbate2002}. In particular the O-$K$ pre-edge fine structure can provide valuable information about Ni-O hybridized states and the associated electronic structure, for instance in the cases of NiO and perovskite $RE$NiO$_3$ \cite{Kuiper1989,Abbate2002,Bisogni2016}. In both materials, Ni 3$d$-orbitals strongly hybridize with oxygen ligands, giving rise to a pre-peak in the absorption spectra near the O $K$-edge (Fig.~\ref{electronic}A). Due to different relative energy scales between the charge-transfer energy $\Delta$ and the Coulomb interaction $U$, according to the Zaanen–Sawatzky–Allen (ZSA) scheme \cite{Zaanen1985}, the former material falls into the regime of charge-transfer insulators, whereas the latter is a negative charge-transfer compound. In contrast, the O $K$-edge absorption spectra of the IL nickelates LaNiO$_2$ and NdNiO$_2$ lack a prominent pre-edge peak (Fig.~\ref{electronic}A), suggesting a substantially weaker effective mixing between oxygen and the unoccupied 3$d$ states of the upper Hubbard band (UHB) of the Ni$^{1+}$ cations \cite{Hepting2020}. In the case of cuprates, a prominent pre-peak feature is present in O $K$-edge absorption spectra \cite{Chen1991}. This is known to originate from the charge-transfer nature of these materials, with $\Delta$ smaller than $U$, and O 2$p$ states mixed with both the lower Hubbard band (LHB) and the UHB of the Cu 3$d_{x^2-y^2}$ states. A sizable pre-peak has also been observed in TL nickelates (Fig.~\ref{electronic}C), indicating mixing between O 2$p$ and the Ni UHB \cite{Zhang2017N, Lin2021}. Upon hole-doping of cuprates, spectral weight shifts from the UHB pre-peak to a lower-energy peak (Fig.~\ref{electronic}C) that is associated with transitions into the doped hole levels \cite{Chen1991,Abbamonte2005}. These hole-states constitute Zhang-Rice singlets (ZRS), which are plaquettes of two doped holes and four O atoms in square planar coordination around a Cu atom, playing the same role as a fully occupied or empty site in an effective single-band Hubbard model \cite{Zhang1988}. Notably, a recent scanning transmission electron microscopy electron-energy loss spectroscopy (STEM-EELS) study suggested that a similar ZRS peak emerges in IL nickelates upon hole-doping \cite{Goodge2021}, although it carries significantly less spectral weight (Fig.~\ref{electronic}C). Overall, there is good evidence for reduced oxygen hybridization in nickelates compared to cuprates, which likely comes from an enhanced $\Delta$ value. Direct and quantivative determination of $\Delta$ and $U$ represents an important issue for future soft x-ray studies.

Further insights into the electronic structure can be obtained from Ni $L$-edge XAS, corresponding to 2$p$-3$d$ multiplet transitions, reflecting the valence configuration of the Ni ions. In the cases of NiO and perovskite $RE$NiO$_3$ (2$p^6$3$d^8$–2$p^5$3$d^9$ and 2$p^6$3$d^8L^n$–2$p^5$3$d^9L^n$ transitions, respectively) distinct multi-peak structures emerge across the $L_3$-edge (Fig.~\ref{electronic}B) \cite{Bisogni2016,Green2016,Hepting2020}. Conversely, the line shapes of the IL nickelates LaNiO$_2$ and especially of NdNiO$_2$ (Fig.~\ref{electronic}B) resemble rather the single-peak XAS spectrum of IL cuprates with only one possible final XAS state (2$p^6$3$d^9$-2$p^5$3$d^{10}$ transition). In more detail, the $L_3$-edge XAS of LaNiO$_2$ and NdNiO$_2$ is dominated by a main peak A (Fig.~\ref{electronic}B), while LaNiO$_2$ shows an additional minor low energy shoulder A' at slightly lower energies.\footnote{The A' feature in XAS is only visible in LaNiO$_2$ films without a SrTiO$_3$ (STO) capping layer. Based on our recent measurements on LaNiO$_2$ films with a STO capping layer, the A' feature, which arises from the resonance of the $\sim$ 0.6 eV feature in the RIXS map, coincides with the the main XAS peak and becomes invisible. In other words, the XAS of La- and Nd-based infinite layer nickelates (with a STO capping layer) are essentially the same.} Figure.~\ref{electronic}D displays the RIXS intensity map of NdNiO$_2$ as a function of the incident photon energy and Fig.~\ref{electronic}E shows RIXS spectra for selected incident energies. Importantly, the RIXS spectra of LaNiO$_2$ and NdNiO$_2$ exhibit a distinct feature around 0.6 eV energy loss (Fig.~\ref{electronic}D,E), which is visible in the RIXS spectra with incident energies coinciding with the XAS peak A (A'). Furthermore, this feature emerges exclusively in the IL compounds and not the perovskite nickelate LaNiO$_3$ (Fig.~\ref{electronic}E). Using exact diagonalization, the general XAS and RIXS features can be reproduced (Fig.~\ref{electronic}F) and the 0.6 eV feature can be assigned to the hybridization between the Ni 3$d_{x^2-y^2}$ and La 5$d$ orbitals, involving a charge-transfer from Ni to the $RE$ cation \cite{Hepting2020}. From LDA+$U$ it is found that a Fermi pocket of mainly La 5$d$ character forms near the $\Gamma$ point, which is quite extended and three-dimensional. On the other hand, the Ni 3$d_{x^2-y^2}$ states in the NiO$_2$ planes are quasi-2D and strongly correlated \cite{Hepting2020,Been2021}. Nevertheless, the relevance of the rare-earth 5$d$ bands for the low-energy physics of IL nickelates is still under debate and proposals range from effective single-band to  multi-orbital models, including various Ni 3$d$ and rare-earth 5$d$ as well as interstitial orbitals \cite{Anisimov1999,Lee2004,Nomura2019,Botana2020,Kitatani2020,Wu2020,Zhang2020E,Sakakibara2020,Lechermann2020,Wang2020E,Werner2020,Adhikary2020,Karp2020,Liu2021,Wan2021,Lang2021,Higashi2021}. Hence, future experiments probing the Fermi surface topology, such as angle resolved photoemission (ARPES) and quantum oscillation measurements, are highly desirable.  

\section{Magnetic correlations}
\subsection{Magnetic excitations in infinite-layer nickelates}

Despite of the involvement of rare-earth 5$d$ states, the fact that the electronic structures of the Ni 3$d$ states resemble a cuprate-like 3$d^9$ system raises a curious question: whether the Mott-physics, a key ingredient in the cuprate phenomenology \cite{Phillips2006}, also play an important role in sculpting the electronic structures in IL nickelates. Since a strong AFM interaction is a consequence of Mott physics due to strong onsite Coulomb interaction, information about the magnetic structures in IL nickelate is imperative to gain further insight into this issue. Early investigations of bulk polycrystalline LaNiO$_2$ and NdNiO$_2$ found no evidence of AFM order \cite{Hayward1999,Hayward2003}, which appeared to suggest a significantly weaker magnetic interaction than in cuprates. On a different ground, theories have been debating the energy scale of magnetic interactions in the IL nickelates. Some theories predict a small AFM interaction ($\sim$ an order of magnitude smaller than that of cuprates) because of the larger charge transfer energy $\Delta$ \cite{Jiang2020,Zhang2020S,Hu2019,Liu2020}. Conversely, other theories argue that the magnetic interactions are comparable to those in cuprates \cite{Been2021, Katukuri2020, Zhang2020S}. Experimental information about magnetic excitations is crucial to clarify this important issue.

Recently, magnetic excitations in Nd$_{1-x}$Sr$_x$NiO$_2$ have been revealed using RIXS at the Ni $L_3$-edge \cite{Lu2021}. As shown in Fig. \ref{magnetic}A, a branch of dispersive magnetic excitations has been observed in NdNiO$_2$, whose energy-momentum dispersion resembles the spin wave excitations of AFM coupled spins in a square lattice. Importantly, the bandwidth of the magnetic excitations is approximately 200 meV, corresponding to a nearest neighbor spin interaction $J_1 \sim$ 65 meV. This is about half of the $J_1$ in cuprate superconductors and similar to that in the TL nickelates \cite{Lin2021}, but is notably higher than in the stripe-ordered single-layer ($n = 1$) nickelates \cite{Sachan1995,Sugai1990,Fabbris2017}, perovskite nickelates ($n=\infty$) \cite{Lu2018}, and cubic NiO \cite{Hutching1972,Ghiringhelli2009}. Therefore, the observation of the high energy scale of $J_1$ in IL nickelates confirms the presence of a strong onsite Coulomb interaction, indicating that the strong correlation effect associated with the Mott-physics is likely also at play in the nickelate superconductors. Notably, distinct from the sharp magnetic modes observed in undoped cuprates, the magnetic excitations in the undoped parent compound of IL nickelates are damped, which is likely due to the coupling to the metallic Nd 5$d$ states.

Upon hole doping, the magnetic excitations become less dispersive as a function of momentum and significantly damped. By fitting the spectrum to a damped harmonic oscillator function, it is found that mode energies soften accompanied by slightly reduced spectral weight (Fig. \ref{magnetic}B). The observed doping dependence is consistent with spin dilution in a Mott insulator. This is in fact different from those observed in cuprates, in which the mode energy and spectral weight do not decrease with increasing doping \cite{Peng2018}. The doping dependence of the magnetic excitations in cuprates has been attributed to the longer-range charge dynamics emergent with increasing hole doping, for example, the three site terms in a Hubbard model \cite{Jia2014,Bala1995}. Such dynamics appear to be less prominent in the doped IL nickelates, likely due to the larger charge transfer energy $\Delta$ and the presence of the rare-earth 5$d$ metallic state, calling for further investigation. 

We note the next nearest-neighbour exchange interaction $J_2$ extracted from the magnetic excitations dispersion possesses an opposite sign to the nearest neighbor $J_1$, which should favor the formation of AFM ordering at (0.5, 0.5). Unfortunately, RIXS at the Ni $L_3$-edge cannot reach (0.5, 0.5) due to insufficient momentum transfer of the photons, preventing a direct scrutinization on the putative AFM order. Notably, recent susceptibility measurement on bulk powder samples indicated spin glass behaviors, but signatures of an AFM phase transition were still not observed. Thus, it would be interesting to investigating why IL nickelates are a failed AFM. However, one should be cautious about the difference between bulk and thin film samples, as well as the disorders in both types of materials, which were significantly reduced over time along with the optimization of material synthesis protocols. 

Interestingly, the IL nickelates add one more case in which the magnetic correlations are in proximity to superconductivity in the phase diagram, similarly to a number of unconventional superconductors, such as cuprates, iron-based superconductors, and heavy fermion superconductors \cite{Scalapino2012}. It might be tempting to attribute magnetic fluctuations as a candidate mechanism of superconductivity. However, among these superconducting compounds, including the nickelate superconductors, there appears no clear correlation between the energy scale of the magnetic excitation and the superconducting transition temperature, casting doubt on this notion. In any case, the relationship between magnetic fluctuations and the superconductivity remains an important issue in nickelate superconductors.  

\subsection{Magnetic excitations in trilayer nickelates}

\begin{figure*}[h!]
\begin{center}
\includegraphics[width=17cm]{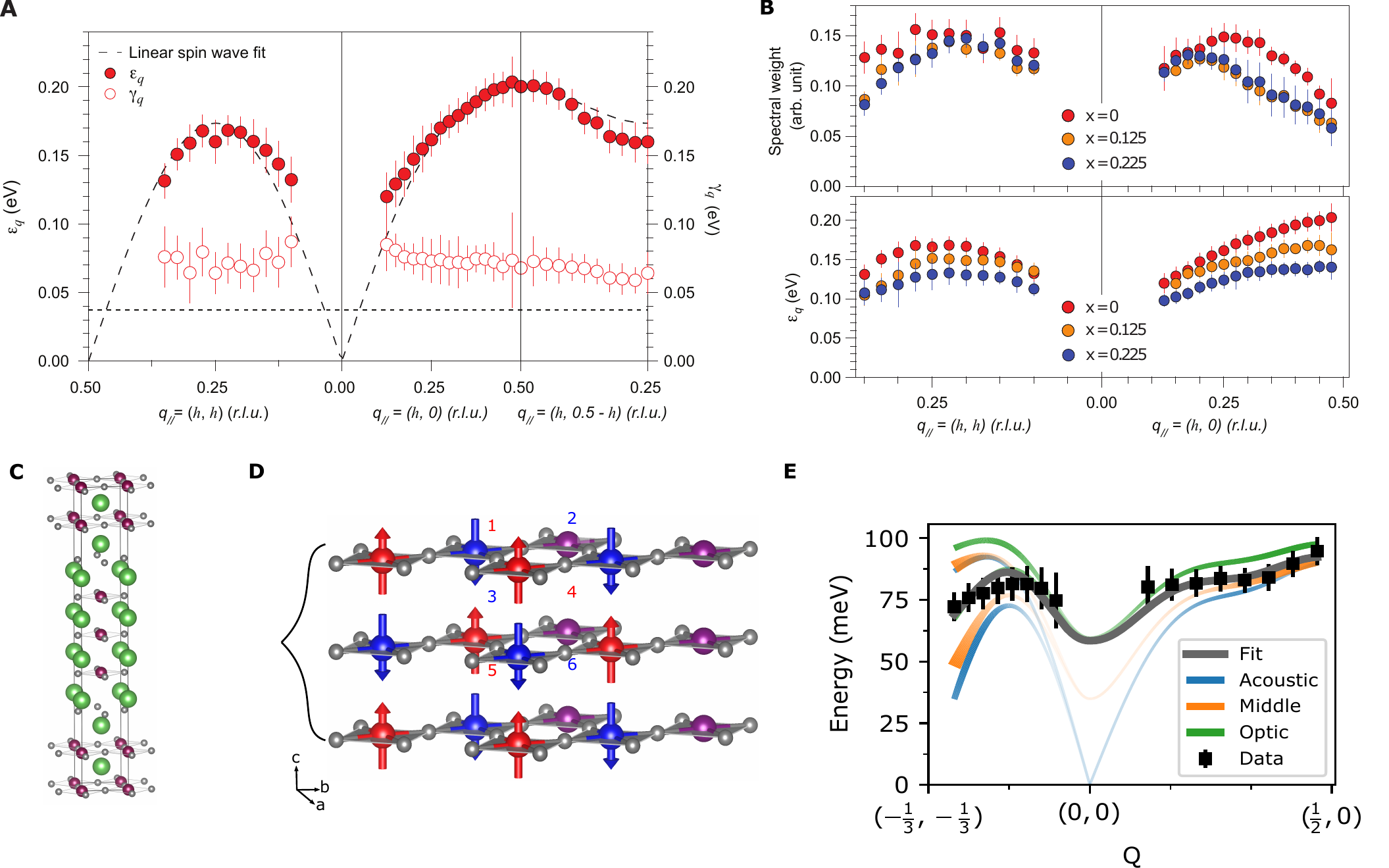}
\end{center}
\caption{Magnetic excitations in nickelates. \textbf{(A)} Magnetic excitations in NdNiO$_2$. Solid and open symbols are mode energy and the damping parameters, respectively, extracted from fitting RIXS spectra to a damped harmonic oscillators (DHO). The dashed curve is the linear spin wave fit to the extracted energy-momentum dispersion. The horizontal dashed line represents the energy resolution of the RIXS measurement. \textbf{(B)} Doping dependence of the spectral weight (upper) and mode energy (lower) deduced from the DHO fitting. \textbf{(C)} The structural unit cell of La$_4$Ni$_3$O$_8$ with Ni/O/La atoms shown as purple/gray/green spheres. \textbf{(D)} The electronically active TL nickel-oxide plane structures in La$_4$Ni$_3$O$_8$, showing the diagonal stripe-ordered state \cite{Zhang2016}. Ni sites with additional hole character are in purple, whereas spinful Ni up(down) sites are depicted in red(blue). \textbf{(E)} Measured magnetic excitations in La$_4$Ni$_3$O$_8$. Black squares are the extracted energies of the magnetic excitations. The dark gray line is the fit to the experimental dispersion, which is composed of three modes plotted in blue, orange, and green, respectively. The doubling of the modes from $(-\frac{1}{3}, -\frac{1}{3})$ to $(0, 0)$ arises from magnetic twinning and the line thickness reflects the predicted intensity of the modes.  Panels adapted from Refs.~\cite{Lu2021, Lin2021}.}
\label{magnetic}
\end{figure*}

In parallel with the measurement of magnetic excitations in the IL material Nd$_{1-x}$Sr$_x$NiO$_2$, magnetic excitations were also measured in the TL materials La$_4$Ni$_3$O$_8$ and Pr$_4$Ni$_3$O$_8$  \cite{Lin2021}. The crystal structure of La$_4$Ni$_3$O$_8$ is shown in Fig.~\ref{magnetic}C. This material has some features that indicate that it might be especially promising as a cuprate analog. The rock salt $RE$-O layers present in its structure make it more two-dimensional than IL compounds and the rare-earth orbitals that are populated in IL are predicted to have a less significant role in TL systems \cite{Zhang2017N, Poltavets2010, Botana2016}. As explained in the introduction, this compound is naturally self-doped and has a nominal hole concentration of 1/3. A disadvantage of the La$_4$Ni$_3$O$_8$ series is that they have, to date, proven difficult to chemically dope. Like cuprates, and some other complex oxides, La$_4$Ni$_3$O$_8$ has charge and spin order \cite{Zhang2016, Zhang2019}. This structure, illustrated in Fig.~\ref{magnetic}D, features diagonal rows of Ni sites with enhanced hole character and neighboring diagonal stripes of up and down spin-ordered sites with reduced hole character. The overall magnetic dispersion, measured with Ni $L_3$ edge RIXS, is plotted in Fig.~\ref{magnetic}E and features a bandwidth of $\sim80$~meV with a downturn near $(-\frac{1}{3}, -\frac{1}{3})$, which is the charge and spin stripe-ordering wavevector. La$_4$Ni$_3$O$_8$ was modeled by solving a Heisenberg Hamiltonian which accounts for stripes, and assumes complete charge disproportionation into $d^9$ and $d^8$ sites, similar to prior studies of other stripe-ordered cuprates and nickelates \cite{Carlson2004spin, Miao2017} . This includes $J_1$, which connects sites 1 and 2 in Fig.~\ref{magnetic}D, $J_3$, which couples atom 2, through the purple doped site, to atom 1 in the next unit cell and $J_z$, which reflects interactions along the $c$-axis, for example, site 1 to site 3. The diagonal $J_2$ interaction that was included in the analysis of NdNiO$_2$ is not expected to be important here, as it couples to the sites with an enhanced hole character and which would be spinless when hosting an extra hole ($S=0$) \cite{Zhang2017N}. Solving this Hamiltonian in the spin-wave approximation yields three modes. Since these modes could not be resolved separately, the RIXS intensity of these modes was computed and summed to predict the intensity of the magnetic feature in RIXS. Values of $J_1 = 69(4)$~meV and $J_3=17(4)$~meV were obtained by fitting a Hamiltonian of this type.\footnote{The notation used here was been modified from the original work of Ref.~\cite{Lin2021} to facilitate comparison with Ref.~\cite{Lu2021}} Similar to the Nd$_{1-x}$Sr$_x$NiO$_2$ case, the effects of $J_z$ were too small to be constrained by the experiment, so the theoretical value of $J_z= 13.6$~meV was used. Very similar dispersions were found in Pr$_4$Ni$_3$O$_8$, which may imply that dynamical stripes exist in this compound even though long range stripe order has not been detected, as has been suggested independently in muon spin rotation studies \cite{Huangfu2020}.

The leading value of $J_1=69(4)$~meV in La$_4$Ni$_3$O$_8$ is strikingly close the 65(1)~meV value obtained for NdNiO$_2$. It should be noted that this similar value arises from a much smaller magnetic bandwidth, as within the stripe-ordered state, each Ni will have only two magnetic neighbors. An approximate extrapolation of the magnetic dispersion of Nd$_{1-x}$Sr$_x$NiO$_2$ to a doping of $x\sim1/3$ implies that it would have a bandwidth comparable to La$_4$Ni$_3$O$_8$ at this doping; yet, whether stripe order or fluctuations exist in the IL nickelates, like those found in the TL nickelates, remains an important open question. Overall, this suggests that the local correlated physics in these reduced RP cousins is very similar provided they are compared at the same effective doping, although their precise low-energy ground states might be more different. 

\section{Conclusion}

In summary, soft x-ray spectroscopic studies have provided valuable insights into the physics of RP-phase and RP-derived nickelates. Nevertheless, for low-valence nickelates there is still limited consensus on the essential ingredients of their electronic structure. Along these lines, we anticipate that advances in sample synthesis and the application of complementary experimental techniques, including ARPES and quantum oscillation measurements, will be helpful. Moreover, the role of disorder and capping layers, as well as the the apparent differences between film and bulk samples need further clarification. These insights could point the way towards improved low-valance nickelate superconductors, including multilayer systems \cite{Ortiz2021}. Finally, a pertinent question is whether  suitable sample preparation allows to realize other cuprate-typical ground states in nickelates, such as antiferromagnetism, pseudogap, as well as nematic, charge, and spin orders.

\section*{Conflict of Interest Statement}

The authors declare that the research was conducted in the absence of any commercial or financial relationships that could be construed as a potential conflict of interest.

\section*{Author Contributions}

The manuscript was written by MH, MPMD, and WSL.

\section*{Funding}

Work at SLAC National Lab was supported by U.S.\ Department of Energy, Office of Science, Office of Basic Energy Sciences under Contract No.~DE-AC02-76SF00515. Work at Brookhaven National Laboratory was supported by the U.S.\ Department of Energy, Office of Science, Office of Basic Energy Sciences. The use of resources at the SIX beamline of the National Synchrotron Light Source II, a U.S.\ Department of Energy (DOE) Office of Science User Facility operated for the DOE Office of Science by Brookhaven National Laboratory under Contract No.~DE-SC0012704, is acknowledged.

\section*{Acknowledgments}

The authors thank K.~F{\"u}rsich for useful discussions and comments.

\bibliographystyle{frontiersinHLTH_FPHY}
\bibliography{bibliography}

\begin{thebibliography}{108}
\expandafter\ifx\csname natexlab\endcsname\relax\def\natexlab#1{#1}\fi
\expandafter\ifx\csname urlstyle\endcsname\relax
  \expandafter\ifx\csname doi\endcsname\relax
  \def\doi#1{doi:\discretionary{}{}{}#1}\fi \else
  \expandafter\ifx\csname doi\endcsname\relax
  \def\doi{doi:\discretionary{}{}{}\begingroup \urlstyle{rm}\Url}\fi \fi
\expandafter\ifx\csname selectlanguage\endcsname\relax
  \def\selectlanguage#1{}\fi

\bibitem[{Keimer et~al.(2015)Keimer, Kivelson, Norman, Uchida, and
  Zaanen}]{Keimer2015}
Keimer B, Kivelson SA, Norman MR, Uchida S, Zaanen J.
\newblock From quantum matter to high-temperature superconductivity in copper
  oxides.
\newblock {\em Nature\/} {\bf 518} (2015) 179--186.
\newblock \doi{10.1038/nature14165}.

\bibitem[{Lee et~al.(2006)Lee, Nagaosa, and Wen}]{Lee2006}
Lee PA, Nagaosa N, Wen XG.
\newblock Doping a {Mott} insulator: Physics of high-temperature
  superconductivity.
\newblock {\em Rev. Mod. Phys.\/} {\bf 78} (2006) 17--85.
\newblock \doi{10.1103/RevModPhys.78.17}.

\bibitem[{Armitage et~al.(2010)Armitage, Fournier, and Greene}]{Armitage2010}
Armitage NP, Fournier P, Greene RL.
\newblock Progress and perspectives on electron-doped cuprates.
\newblock {\em Rev. Mod. Phys.\/} {\bf 82} (2010) 2421--2487.
\newblock \doi{10.1103/RevModPhys.82.2421}.

\bibitem[{Damascelli et~al.(2003)Damascelli, Hussain, and
  Shen}]{Damascelli2003}
Damascelli A, Hussain Z, Shen ZX.
\newblock Angle-resolved photoemission studies of the cuprate superconductors.
\newblock {\em Rev. Mod. Phys.\/} {\bf 75} (2003) 473--541.
\newblock \doi{10.1103/RevModPhys.75.473}.

\bibitem[{Fink et~al.(2001)Fink, Knupfer, Atzkern, and Golden}]{Fink2001}
Fink J, Knupfer M, Atzkern S, Golden M.
\newblock Electronic correlations in solids, studied using electron energy-loss
  spectroscopy.
\newblock {\em J. Electron Spectros. Relat. Phenomena\/} {\bf 117-118} (2001)
  287--309.
\newblock \doi{10.1016/S0368-2048(01)00254-7}.

\bibitem[{Fischer et~al.(2007)Fischer, Kugler, Maggio-Aprile, Berthod, and
  Renner}]{Fischer2007}
Fischer O, Kugler M, Maggio-Aprile I, Berthod C, Renner C.
\newblock Scanning tunneling spectroscopy of high-temperature superconductors.
\newblock {\em Rev. Mod. Phys.\/} {\bf 79} (2007) 353--419.
\newblock \doi{10.1103/RevModPhys.79.353}.

\bibitem[{Basov and Timusk(2005)}]{Basov2005}
Basov DN, Timusk T.
\newblock Electrodynamics of high-${T}_{c}$ superconductors.
\newblock {\em Rev. Mod. Phys.\/} {\bf 77} (2005) 721--779.
\newblock \doi{10.1103/RevModPhys.77.721}.

\bibitem[{Devereaux and Hackl(2007)}]{Devereaux2007}
Devereaux TP, Hackl R.
\newblock Inelastic light scattering from correlated electrons.
\newblock {\em Rev. Mod. Phys.\/} {\bf 79} (2007) 175--233.
\newblock \doi{10.1103/RevModPhys.79.175}.

\bibitem[{Fujita et~al.(2012)Fujita, Hiraka, Matsuda, Matsuura, M.~Tranquada,
  Wakimoto et~al.}]{Fujita2012}
Fujita M, Hiraka H, Matsuda M, Matsuura M, M~Tranquada J, Wakimoto S, et~al.
\newblock Progress in neutron scattering studies of spin excitations in
  high-{$T_c$} cuprates.
\newblock {\em J. Phys. Soc. Japan\/} {\bf 81} (2012) 011007.
\newblock \doi{10.1143/JPSJ.81.011007}.

\bibitem[{Miao et~al.(2018)Miao, Ishikawa, Heid, Le~Tacon, Fabbris, Meyers
  et~al.}]{Miao2018}
Miao H, Ishikawa D, Heid R, Le~Tacon M, Fabbris G, Meyers D, et~al.
\newblock Incommensurate phonon anomaly and the nature of charge density waves
  in cuprates.
\newblock {\em Phys. Rev. X\/} {\bf 8} (2018) 011008.
\newblock \doi{10.1103/PhysRevX.8.011008}.

\bibitem[{Ament et~al.(2011)Ament, van Veenendaal, Devereaux, Hill, and van~den
  Brink}]{Ament2011}
Ament LJP, van Veenendaal M, Devereaux TP, Hill JP, van~den Brink J.
\newblock Resonant inelastic x-ray scattering studies of elementary
  excitations.
\newblock {\em Rev. Mod. Phys.\/} {\bf 83} (2011) 705--767.
\newblock \doi{10.1103/RevModPhys.83.705}.

\bibitem[{Le~Tacon et~al.(2011)Le~Tacon, Ghiringhelli, Chaloupka, Sala, Hinkov,
  Haverkort et~al.}]{LeTacon2011}
Le~Tacon M, Ghiringhelli G, Chaloupka J, Sala MM, Hinkov V, Haverkort M, et~al.
\newblock Intense paramagnon excitations in a large family of high-temperature
  superconductors.
\newblock {\em Nat. Phys.\/} {\bf 7} (2011) 725--730.
\newblock \doi{10.1038/NPHYS2041}.

\bibitem[{Dean et~al.(2013)Dean, Dellea, Springell, Yakhou-Harris, Kummer,
  Brookes et~al.}]{Dean2013}
Dean M, Dellea G, Springell R, Yakhou-Harris F, Kummer K, Brookes N, et~al.
\newblock Persistence of magnetic excitations in {La$_{2-x}$Sr$_x$CuO$_4$} from
  the undoped insulator to the heavily overdoped non-superconducting metal.
\newblock {\em Nat. Mater.\/} {\bf 12} (2013) 1019--1023.
\newblock \doi{10.1038/NMAT3723}.

\bibitem[{Hepting et~al.(2018{\natexlab{a}})Hepting, Chaix, Huang, Fumagalli,
  Peng, Moritz et~al.}]{Hepting2018}
Hepting M, Chaix L, Huang E, Fumagalli R, Peng Y, Moritz B, et~al.
\newblock Three-dimensional collective charge excitations in electron-doped
  copper oxide superconductors.
\newblock {\em Nature\/} {\bf 563} (2018{\natexlab{a}}) 374--378.
\newblock \doi{10.1038/s41586-018-0648-3}.

\bibitem[{{J. Q. Lin, Jie Yuan, Kui Jin, Z. P. Yin, Gang Li, Ke-Jin Zhou,
  Xingye Lu, M. Dantz, Thorsten Schmitt, H. Ding, Haizhong Guo, M. P. M. Dean,
  and X. Liu}(2020)}]{Lin2019}
{J Q Lin, Jie Yuan, Kui Jin, Z P Yin, Gang Li, Ke-Jin Zhou, Xingye Lu, M Dantz,
  Thorsten Schmitt, H Ding, Haizhong Guo, M P M Dean, and X Liu}.
\newblock Doping evolution of the charge excitations and electron correlations
  in electron-doped superconducting {La$_{2-x}$Ce$_x$CuO$_4$}.
\newblock {\em npj Quantum Mater.\/} {\bf 5} (2020) 4.
\newblock \doi{10.1038/s41535-019-0205-9}.

\bibitem[{Nag et~al.(2020)Nag, Zhu, Bejas, Li, Robarts, Yamase
  et~al.}]{Nag2020}
Nag A, Zhu M, Bejas M, Li J, Robarts HC, Yamase H, et~al.
\newblock Detection of acoustic plasmons in hole-doped lanthanum and bismuth
  cuprate superconductors using resonant inelastic x-ray scattering.
\newblock {\em Phys. Rev. Lett.\/} {\bf 125} (2020) 257002.
\newblock \doi{10.1103/PhysRevLett.125.257002}.

\bibitem[{Ghiringhelli et~al.(2012)Ghiringhelli, {Le Tacon}, Minola,
  Blanco-Canosa, Mazzoli, Brookes et~al.}]{Ghiringhelli2012}
Ghiringhelli G, {Le Tacon} M, Minola M, Blanco-Canosa S, Mazzoli C, Brookes NB,
  et~al.
\newblock {Long-Range Incommensurate Charge Fluctuations in
  (Y,Nd)Ba$_{2}$Cu$_{3}$O$_{6+x}$}.
\newblock {\em Science\/} {\bf 337} (2012) 821--825.
\newblock \doi{10.1126/science.1223532}.

\bibitem[{Miao et~al.(2017)Miao, Lorenzana, Seibold, Peng, Amorese,
  Yakhou-Harris et~al.}]{Miao2017}
Miao H, Lorenzana J, Seibold G, Peng YY, Amorese A, Yakhou-Harris F, et~al.
\newblock High-temperature charge density wave correlations in
  {La$_{1.875}$Ba$_{0.125}$CuO$_4$} without spin{\textendash}charge locking.
\newblock {\em PNAS\/} {\bf 114} (2017) 12430--12435.
\newblock \doi{10.1073/pnas.1708549114}.

\bibitem[{Chaix et~al.(2017)Chaix, Ghiringhelli, Peng, Hashimoto, Moritz,
  Kummer et~al.}]{Chaix2017}
Chaix L, Ghiringhelli G, Peng Y, Hashimoto M, Moritz B, Kummer K, et~al.
\newblock Dispersive charge density wave excitations in
  {Bi$_2$Sr$_2$CaCu$_2$O$_{8+\delta}$}.
\newblock {\em Nat. Phys.\/} {\bf 13} (2017) 952--956.
\newblock \doi{10.1038/nphys4157}.

\bibitem[{Miao et~al.(2019)Miao, Fumagalli, Rossi, Lorenzana, Seibold,
  Yakhou-Harris et~al.}]{Miao2019a}
Miao H, Fumagalli R, Rossi M, Lorenzana J, Seibold G, Yakhou-Harris F, et~al.
\newblock Formation of incommensurate charge density waves in cuprates.
\newblock {\em Phys. Rev. X\/} {\bf 9} (2019) 031042.
\newblock \doi{10.1103/PhysRevX.9.031042}.

\bibitem[{Lee et~al.(2021)Lee, Zhou, Hepting, Li, Nag, Walters
  et~al.}]{Lee2021}
Lee WS, Zhou KJ, Hepting M, Li J, Nag A, Walters AC, et~al.
\newblock {Spectroscopic fingerprint of charge order melting driven by quantum
  fluctuations in a cuprate}.
\newblock {\em Nat. Phys.\/} {\bf 17} (2021) 53--57.
\newblock \doi{10.1038/s41567-020-0993-7}.

\bibitem[{Norman(2016)}]{Norman2016}
Norman MR.
\newblock Materials design for new superconductors.
\newblock {\em Rep. Prog. Phys.\/} {\bf 79} (2016) 074502.
\newblock \doi{10.1088/0034-4885/79/7/074502}.

\bibitem[{Greenblatt(1997)}]{Greenblatt1997}
Greenblatt M.
\newblock Ruddlesden-popper {Ln$_{n+1}$Ni$_n$O$_{3n+1}$} nickelates: structure
  and properties.
\newblock {\em Curr. Opin. Solid State Mater. Sci.\/} {\bf 2} (1997) 174--183.
\newblock \doi{10.1016/S1359-0286(97)80062-9}.

\bibitem[{Acrivos et~al.(1994)Acrivos, Lei, Jiang, Nguyen, Metcalf, and
  Honig}]{Acrivos1994}
Acrivos J, Lei M, Jiang C, Nguyen H, Metcalf P, Honig J.
\newblock Paramagnetism, antiferromagnetism, and superconductivity in
  {La$_2$NiO$_4$}.
\newblock {\em J. Solid State Chem.\/} {\bf 111} (1994) 343--348.
\newblock \doi{10.1006/jssc.1994.1237}.

\bibitem[{Sachan et~al.(1995)Sachan, Buttrey, Tranquada, Lorenzo, and
  Shirane}]{Sachan1995}
Sachan V, Buttrey DJ, Tranquada JM, Lorenzo JE, Shirane G.
\newblock Charge and spin ordering in
  {${\mathrm{La}}_{2\mathrm{\ensuremath{-}}\mathit{x}}$${\mathrm{Sr}}_{\mathit{x}}$${\mathrm{NiO}}_{4.00}$
  with x=0.135 and 0.20}.
\newblock {\em Phys. Rev. B\/} {\bf 51} (1995) 12742--12746.
\newblock \doi{10.1103/PhysRevB.51.12742}.

\bibitem[{Sugai et~al.(1990)Sugai, Sato, Kobayashi, Akimitsu, Ito, Takagi
  et~al.}]{Sugai1990}
Sugai S, Sato M, Kobayashi T, Akimitsu J, Ito T, Takagi H, et~al.
\newblock High-energy spin excitations in the insulating phases of
  high-{${\mathit{T}}_{\mathit{c}}$} superconducting cuprates and
  {${\mathrm{La}}_{2}{\mathrm{NiO}}_{4}$}.
\newblock {\em Phys. Rev. B\/} {\bf 42} (1990) 1045--1047.
\newblock \doi{10.1103/PhysRevB.42.1045}.

\bibitem[{Fabbris et~al.(2017)Fabbris, Meyers, Xu, Katukuri, Hozoi, Liu
  et~al.}]{Fabbris2017}
Fabbris G, Meyers D, Xu L, Katukuri VM, Hozoi L, Liu X, et~al.
\newblock {Doping Dependence of Collective Spin and Orbital Excitations in the
  Spin-1 Quantum Antiferromagnet
  ${\mathrm{La}}_{2\ensuremath{-}x}{\mathrm{Sr}}_{x}{\mathrm{NiO}}_{4}$
  Observed by X Rays}.
\newblock {\em Phys. Rev. Lett.\/} {\bf 118} (2017) 156402.
\newblock \doi{10.1103/PhysRevLett.118.156402}.

\bibitem[{Bisogni et~al.(2016)Bisogni, Catalano, Green, Gibert, Scherwitzl,
  Huang et~al.}]{Bisogni2016}
Bisogni V, Catalano S, Green RJ, Gibert M, Scherwitzl R, Huang Y, et~al.
\newblock {Ground-state oxygen holes and the metal–insulator transition in
  the negative charge-transfer rare-earth nickelates}.
\newblock {\em Nat. Commun.\/} {\bf 7} (2016) 13017.
\newblock \doi{10.1038/ncomms13017}.

\bibitem[{Chaloupka and Khaliullin(2008)}]{Chaloupka2008}
Chaloupka Jcv, Khaliullin G.
\newblock Orbital order and possible superconductivity in
  {${\mathrm{LaNiO}}_{3}/{\mathrm{LaMO}}_{3}$} superlattices.
\newblock {\em Phys. Rev. Lett.\/} {\bf 100} (2008) 016404.
\newblock \doi{10.1103/PhysRevLett.100.016404}.

\bibitem[{Hansmann et~al.(2009)Hansmann, Yang, Toschi, Khaliullin, Andersen,
  and Held}]{Hansmann2009}
Hansmann P, Yang X, Toschi A, Khaliullin G, Andersen OK, Held K.
\newblock Turning a nickelate fermi surface into a cupratelike one through
  heterostructuring.
\newblock {\em Phys. Rev. Lett.\/} {\bf 103} (2009) 016401.
\newblock \doi{10.1103/PhysRevLett.103.016401}.

\bibitem[{Middey et~al.(2016)Middey, Chakhalian, Mahadevan, Freeland, Millis,
  and Sarma}]{Middey2016}
Middey S, Chakhalian J, Mahadevan P, Freeland J, Millis A, Sarma D.
\newblock Physics of ultrathin films and heterostructures of rare-earth
  nickelates.
\newblock {\em Annu. Rev. Mater. Res.\/} {\bf 46} (2016) 305--334.
\newblock \doi{10.1146/annurev-matsci-070115-032057}.

\bibitem[{Catalano et~al.(2018)Catalano, Gibert, Fowlie, {\'{I}}{\~{n}}iguez,
  Triscone, and Kreisel}]{Catalano2018}
Catalano S, Gibert M, Fowlie J, {\'{I}}{\~{n}}iguez J, Triscone JM, Kreisel J.
\newblock Rare-earth nickelates {$R$NiO$_3$}: thin films and heterostructures.
\newblock {\em Rep. Prog. Phys.\/} {\bf 81} (2018) 046501.
\newblock \doi{10.1088/1361-6633/aaa37a}.

\bibitem[{Boris et~al.(2011)Boris, Matiks, Benckiser, Frano, Popovich, Hinkov
  et~al.}]{Boris2011}
Boris AV, Matiks Y, Benckiser E, Frano A, Popovich P, Hinkov V, et~al.
\newblock Dimensionality control of electronic phase transitions in
  nickel-oxide superlattices.
\newblock {\em Science\/} {\bf 332} (2011) 937--940.
\newblock \doi{10.1126/science.1202647}.

\bibitem[{Benckiser et~al.(2011)Benckiser, Haverkort, Br{\"{u}}ck, Goering,
  Macke, Fra{\~{n}}{\'{o}} et~al.}]{Benckiser2011}
Benckiser E, Haverkort MW, Br{\"{u}}ck S, Goering E, Macke S, Fra{\~{n}}{\'{o}}
  A, et~al.
\newblock {Orbital reflectometry of oxide heterostructures}.
\newblock {\em Nat. Mater.\/} {\bf 10} (2011) 189--193.
\newblock \doi{10.1038/NMAT2958}.

\bibitem[{Liu et~al.(2011)Liu, Okamoto, van Veenendaal, Kareev, Gray, Ryan
  et~al.}]{Liu2011}
Liu J, Okamoto S, van Veenendaal M, Kareev M, Gray B, Ryan P, et~al.
\newblock Quantum confinement of mott electrons in ultrathin
  {LaNiO${}_{3}$/LaAlO${}_{3}$} superlattices.
\newblock {\em Phys. Rev. B\/} {\bf 83} (2011) 161102.
\newblock \doi{10.1103/PhysRevB.83.161102}.

\bibitem[{Frano et~al.(2013)Frano, Schierle, Haverkort, Lu, Wu, Blanco-Canosa
  et~al.}]{Frano2013}
Frano A, Schierle E, Haverkort MW, Lu Y, Wu M, Blanco-Canosa S, et~al.
\newblock Orbital control of noncollinear magnetic order in nickel oxide
  heterostructures.
\newblock {\em Phys. Rev. Lett.\/} {\bf 111} (2013) 106804.
\newblock \doi{10.1103/PhysRevLett.111.106804}.

\bibitem[{Hepting et~al.(2014)Hepting, Minola, Frano, Cristiani, Logvenov,
  Schierle et~al.}]{Hepting2014}
Hepting M, Minola M, Frano A, Cristiani G, Logvenov G, Schierle E, et~al.
\newblock Tunable charge and spin order in {${\mathrm{PrNiO}}_{3}$} thin films
  and superlattices.
\newblock {\em Phys. Rev. Lett.\/} {\bf 113} (2014) 227206.
\newblock \doi{10.1103/PhysRevLett.113.227206}.

\bibitem[{Gibert et~al.(2015)Gibert, Viret, Torres-Pardo, Piamonteze, Zubko,
  Jaouen et~al.}]{Gibert2015}
Gibert M, Viret M, Torres-Pardo A, Piamonteze C, Zubko P, Jaouen N, et~al.
\newblock Interfacial control of magnetic properties at
  {${\mathrm{LaMnO}}_{3}$/${\mathrm{LaNiO}}_{3}$} interfaces.
\newblock {\em Nano Lett.\/} {\bf 15} (2015) 7355--7361.
\newblock \doi{10.1021/acs.nanolett.5b02720}.

\bibitem[{Disa et~al.(2017)Disa, Georgescu, Hart, Kumah, Shafer, Arenholz
  et~al.}]{Disa2017}
Disa AS, Georgescu AB, Hart JL, Kumah DP, Shafer P, Arenholz E, et~al.
\newblock Control of hidden ground-state order in
  {$\mathrm{NdNi}{\mathrm{O}}_{3}$} superlattices.
\newblock {\em Phys. Rev. Materials\/} {\bf 1} (2017) 024410.
\newblock \doi{10.1103/PhysRevMaterials.1.024410}.

\bibitem[{Hepting et~al.(2018{\natexlab{b}})Hepting, Green, Zhong, Bluschke,
  Suyolcu, Macke et~al.}]{Hepting2018a}
Hepting M, Green RJ, Zhong Z, Bluschke M, Suyolcu YE, Macke S, et~al.
\newblock {Complex magnetic order in nickelate slabs}.
\newblock {\em Nat. Phys.\/} {\bf 14} (2018{\natexlab{b}}) 1097--1102.
\newblock \doi{10.1038/s41567-018-0218-5}.

\bibitem[{F\"ursich et~al.(2019)F\"ursich, Lu, Betto, Bluschke, Porras,
  Schierle et~al.}]{Fursich2019}
F\"ursich K, Lu Y, Betto D, Bluschke M, Porras J, Schierle E, et~al.
\newblock Resonant inelastic x-ray scattering study of bond order and spin
  excitations in nickelate thin-film structures.
\newblock {\em Phys. Rev. B\/} {\bf 99} (2019) 165124.
\newblock \doi{10.1103/PhysRevB.99.165124}.

\bibitem[{Chakhalian et~al.(2011)Chakhalian, Rondinelli, Liu, Gray, Kareev,
  Moon et~al.}]{Chakhalian2011}
Chakhalian J, Rondinelli JM, Liu J, Gray BA, Kareev M, Moon EJ, et~al.
\newblock Asymmetric orbital-lattice interactions in ultrathin correlated oxide
  films.
\newblock {\em Phys. Rev. Lett.\/} {\bf 107} (2011) 116805.
\newblock \doi{10.1103/PhysRevLett.107.116805}.

\bibitem[{Wu et~al.(2013)Wu, Benckiser, Haverkort, Frano, Lu, Nwankwo
  et~al.}]{Wu2013}
Wu M, Benckiser E, Haverkort MW, Frano A, Lu Y, Nwankwo U, et~al.
\newblock Strain and composition dependence of orbital polarization in nickel
  oxide superlattices.
\newblock {\em Phys. Rev. B\/} {\bf 88} (2013) 125124.
\newblock \doi{10.1103/PhysRevB.88.125124}.

\bibitem[{Disa et~al.(2015)Disa, Kumah, Malashevich, Chen, Arena, Specht
  et~al.}]{Disa2015}
Disa AS, Kumah DP, Malashevich A, Chen H, Arena DA, Specht ED, et~al.
\newblock Orbital engineering in symmetry-breaking polar heterostructures.
\newblock {\em Phys. Rev. Lett.\/} {\bf 114} (2015) 026801.
\newblock \doi{10.1103/PhysRevLett.114.026801}.

\bibitem[{Han et~al.(2011)Han, Wang, Marianetti, and Millis}]{Han2011}
Han MJ, Wang X, Marianetti CA, Millis AJ.
\newblock Dynamical mean-field theory of nickelate superlattices.
\newblock {\em Phys. Rev. Lett.\/} {\bf 107} (2011) 206804.
\newblock \doi{10.1103/PhysRevLett.107.206804}.

\bibitem[{Fabbris et~al.(2016)Fabbris, Meyers, Okamoto, Pelliciari, Disa, Huang
  et~al.}]{Fabbris2016}
Fabbris G, Meyers D, Okamoto J, Pelliciari J, Disa AS, Huang Y, et~al.
\newblock Orbital engineering in nickelate heterostructures driven by
  anisotropic oxygen hybridization rather than orbital energy levels.
\newblock {\em Phys. Rev. Lett.\/} {\bf 117} (2016) 147401.
\newblock \doi{10.1103/PhysRevLett.117.147401}.

\bibitem[{Zhang et~al.(2016)Zhang, Chen, Phelan, Zheng, Norman, and
  Mitchell}]{Zhang2016}
Zhang J, Chen YS, Phelan D, Zheng H, Norman MR, Mitchell JF.
\newblock Stacked charge stripes in the quasi-{2D} trilayer nickelate
  {La$_4$Ni$_3$O$_8$}.
\newblock {\em PNAS\/} {\bf 113} (2016) 8945--8950.
\newblock \doi{10.1073/pnas.1606637113}.

\bibitem[{Zhang et~al.(2017)Zhang, Botana, Freeland, Phelan, Zheng, Pardo
  et~al.}]{Zhang2017N}
Zhang J, Botana AS, Freeland JW, Phelan D, Zheng H, Pardo V, et~al.
\newblock Large orbital polarization in a metallic square-planar nickelate.
\newblock {\em Nat. Phys.\/} {\bf 13} (2017) 864--869.
\newblock \doi{10.1038/nphys4149}.

\bibitem[{Lin et~al.(2021)Lin, Villar~Arribi, Fabbris, Botana, Meyers, Miao
  et~al.}]{Lin2021}
Lin JQ, Villar~Arribi P, Fabbris G, Botana AS, Meyers D, Miao H, et~al.
\newblock Strong superexchange in a ${d}^{9\ensuremath{-}\ensuremath{\delta}}$
  nickelate revealed by resonant inelastic x-ray scattering.
\newblock {\em Phys. Rev. Lett.\/} {\bf 126} (2021) 087001.
\newblock \doi{10.1103/PhysRevLett.126.087001}.

\bibitem[{Botana et~al.(2017)Botana, Pardo, and Norman}]{Botana2017}
Botana AS, Pardo V, Norman MR.
\newblock Electron doped layered nickelates: Spanning the phase diagram of the
  cuprates.
\newblock {\em Phys. Rev. Materials\/} {\bf 1} (2017) 021801.
\newblock \doi{10.1103/PhysRevMaterials.1.021801}.

\bibitem[{Nica et~al.(2020)Nica, Krishna, Yu, Si, Botana, and Erten}]{Nica2020}
Nica EM, Krishna J, Yu R, Si Q, Botana AS, Erten O.
\newblock Theoretical investigation of superconductivity in trilayer
  square-planar nickelates.
\newblock {\em Phys. Rev. B\/} {\bf 102} (2020) 020504.
\newblock \doi{10.1103/PhysRevB.102.020504}.

\bibitem[{Li et~al.(2019)Li, Lee, Wang, Osada, Crossley, Lee et~al.}]{Li2019}
Li D, Lee K, Wang BY, Osada M, Crossley S, Lee HR, et~al.
\newblock Superconductivity in an infinite-layer nickelate.
\newblock {\em Nature\/} {\bf 572} (2019) 624--627.
\newblock \doi{10.1038/s41586-019-1496-5}.

\bibitem[{Zeng et~al.(2020)Zeng, Tang, Yin, Li, Li, Huang et~al.}]{Zeng2020}
Zeng S, Tang CS, Yin X, Li C, Li M, Huang Z, et~al.
\newblock Phase diagram and superconducting dome of infinite-layer
  {Nd$_{1-x}$Sr$_x$NiO$_2$} thin films.
\newblock {\em Phys. Rev. Lett.\/} {\bf 125} (2020) 147003.
\newblock \doi{10.1103/physrevlett.125.147003}.

\bibitem[{Li et~al.(2020)Li, Wang, Lee, Harvey, Osada, Goodge et~al.}]{Li2020a}
Li D, Wang BY, Lee K, Harvey SP, Osada M, Goodge BH, et~al.
\newblock Superconducting dome in
  {${\mathrm{Nd}}_{1\ensuremath{-}x}{\mathrm{Sr}}_{x}{\mathrm{NiO}}_{2}$}
  infinite layer films.
\newblock {\em Phys. Rev. Lett.\/} {\bf 125} (2020) 027001.
\newblock \doi{10.1103/PhysRevLett.125.027001}.

\bibitem[{Lee et~al.(2020)Lee, Goodge, Li, Osada, Wang, Cui et~al.}]{Lee2020}
Lee K, Goodge BH, Li D, Osada M, Wang BY, Cui Y, et~al.
\newblock Aspects of the synthesis of thin film superconducting infinite-layer
  nickelates.
\newblock {\em {APL} Mater.\/} {\bf 8} (2020) 041107.
\newblock \doi{10.1063/5.0005103}.

\bibitem[{Osada et~al.(2020)Osada, Wang, Goodge, Lee, Yoon, Sakuma
  et~al.}]{Osada2020}
Osada M, Wang BY, Goodge BH, Lee K, Yoon H, Sakuma K, et~al.
\newblock A superconducting praseodymium nickelate with infinite layer
  structure.
\newblock {\em Nano Lett.\/} {\bf 20} (2020) 5735--5740.
\newblock \doi{10.1021/acs.nanolett.0c01392}.

\bibitem[{Li et~al.(2021)Li, Sun, Yang, Cai, Guo, Gu et~al.}]{Li2021G}
Li Y, Sun W, Yang J, Cai X, Guo W, Gu Z, et~al.
\newblock Impact of cation stoichiometry on the crystalline structure and
  superconductivity in nickelates.
\newblock {\em Front. Phys.\/} {\bf 9} (2021) 443.
\newblock \doi{10.3389/fphy.2021.719534}.

\bibitem[{Osada et~al.(2021)Osada, Wang, Goodge, Harvey, Lee, Li
  et~al.}]{Osada2021}
Osada M, Wang BY, Goodge BH, Harvey SP, Lee K, Li D, et~al.
\newblock Nickelate superconductivity without rare-earth magnetism:
  {(La,Sr)NiO$_2$}.
\newblock {\em Adv. Mater.\/} {\bf n/a} (2021) 2104083.
\newblock \doi{10.1002/adma.202104083}.

\bibitem[{Hayward et~al.(1999)Hayward, Green, Rosseinsky, and
  Sloan}]{Hayward1999}
Hayward MA, Green MA, Rosseinsky MJ, Sloan J.
\newblock Sodium hydride as a powerful reducing agent for topotactic oxide
  deintercalation:~ synthesis and characterization of the nickel(i) oxide
  {LaNiO$_2$}.
\newblock {\em J. Am. Chem. Soc.\/} {\bf 121} (1999) 8843--8854.
\newblock \doi{10.1021/ja991573i}.

\bibitem[{Hayward and Rosseinsky(2003)}]{Hayward2003}
Hayward M, Rosseinsky M.
\newblock Synthesis of the infinite layer ni(i) phase {NdNiO$_{2+x}$} by low
  temperature reduction of {NdNiO$_3$} with sodium hydride.
\newblock {\em Solid State Sci.\/} {\bf 5} (2003) 839--850.
\newblock \doi{10.1016/S1293-2558(03)00111-0}.

\bibitem[{Ikeda et~al.(2016)Ikeda, Krockenberger, Irie, Naito, and
  Yamamoto}]{Ikeda2016}
Ikeda A, Krockenberger Y, Irie H, Naito M, Yamamoto H.
\newblock Direct observation of infinite {NiO$_2$} planes in {LaNiO$_2$} films.
\newblock {\em Appl. Phys. Express\/} {\bf 9} (2016) 061101.
\newblock \doi{10.7567/apex.9.061101}.

\bibitem[{Anisimov et~al.(1999)Anisimov, Bukhvalov, and Rice}]{Anisimov1999}
Anisimov VI, Bukhvalov D, Rice TM.
\newblock Electronic structure of possible nickelate analogs to the cuprates.
\newblock {\em Phys. Rev. B\/} {\bf 59} (1999) 7901--7906.
\newblock \doi{10.1103/physrevb.59.7901}.

\bibitem[{Lee and Pickett(2004)}]{Lee2004}
Lee KW, Pickett WE.
\newblock Infinite-layer {LaNiO$_2$}: {Ni$^{1+}$} is not {Cu$^{2+}$}.
\newblock {\em Phys. Rev. B\/} {\bf 70} (2004) 165109.
\newblock \doi{10.1103/physrevb.70.165109}.

\bibitem[{Hepting et~al.(2020)Hepting, Li, Jia, Lu, Paris, Tseng
  et~al.}]{Hepting2020}
Hepting M, Li D, Jia CJ, Lu H, Paris E, Tseng Y, et~al.
\newblock Electronic structure of the parent compound of superconducting
  infinite-layer nickelates.
\newblock {\em Nat. Mater.\/} {\bf 19} (2020) 381--385.
\newblock \doi{10.1038/s41563-019-0585-z}.

\bibitem[{Goodge et~al.(2021)Goodge, Li, Lee, Osada, Wang, Sawatzky
  et~al.}]{Goodge2021}
Goodge BH, Li D, Lee K, Osada M, Wang BY, Sawatzky GA, et~al.
\newblock Doping evolution of the mott{\textendash}hubbard landscape in
  infinite-layer nickelates.
\newblock {\em PNAS\/} {\bf 118} (2021) e2007683118.
\newblock \doi{10.1073/pnas.2007683118}.

\bibitem[{Lu et~al.(2021)Lu, Rossi, Nag, Osada, Li, Lee et~al.}]{Lu2021}
Lu H, Rossi M, Nag A, Osada M, Li DF, Lee K, et~al.
\newblock Magnetic excitations in infinite-layer nickelates.
\newblock {\em Science\/} {\bf 373} (2021) 213--216.
\newblock \doi{10.1126/science.abd7726}.

\bibitem[{Abbate et~al.(2002)Abbate, Zampieri, Prado, Caneiro, Gonzalez-Calbet,
  and Vallet-Regi}]{Abbate2002}
Abbate M, Zampieri G, Prado F, Caneiro A, Gonzalez-Calbet JM, Vallet-Regi M.
\newblock Electronic structure and metal-insulator transition in
  {LaNiO$_{3-\delta}$}.
\newblock {\em Phys. Rev. B\/} {\bf 65} (2002) 155101.
\newblock \doi{10.1103/physrevb.65.155101}.

\bibitem[{Kuiper et~al.(1989)Kuiper, Kruizinga, Ghijsen, Sawatzky, and
  Verweij}]{Kuiper1989}
Kuiper P, Kruizinga G, Ghijsen J, Sawatzky GA, Verweij H.
\newblock Character of holes in
  {${\mathrm{Li}}_{x}{\mathrm{Ni}}_{1\ensuremath{-}x}\mathrm{O}$} and their
  magnetic behavior.
\newblock {\em Phys. Rev. Lett.\/} {\bf 62} (1989) 221--224.
\newblock \doi{10.1103/PhysRevLett.62.221}.

\bibitem[{Zaanen et~al.(1985)Zaanen, Sawatzky, and Allen}]{Zaanen1985}
Zaanen J, Sawatzky GA, Allen JW.
\newblock Band gaps and electronic structure of transition-metal compounds.
\newblock {\em Phys. Rev. Lett.\/} {\bf 55} (1985) 418--421.
\newblock \doi{10.1103/PhysRevLett.55.418}.

\bibitem[{Chen et~al.(1991)Chen, Sette, Ma, Hybertsen, Stechel, Foulkes
  et~al.}]{Chen1991}
Chen CT, Sette F, Ma Y, Hybertsen MS, Stechel EB, Foulkes WMC, et~al.
\newblock Electronic states in
  {${\mathrm{La}}_{2\mathrm{\ensuremath{-}}\mathit{x}}$${\mathrm{Sr}}_{\mathit{x}}$${\mathrm{CuO}}_{4+\mathrm{\ensuremath{\delta}}}$}
  probed by soft-x-ray absorption.
\newblock {\em Phys. Rev. Lett.\/} {\bf 66} (1991) 104--107.
\newblock \doi{10.1103/PhysRevLett.66.104}.

\bibitem[{Abbamonte et~al.(2005)Abbamonte, Rusydi, Smadici, Gu, Sawatzky, and
  Feng}]{Abbamonte2005}
Abbamonte P, Rusydi A, Smadici S, Gu GD, Sawatzky GA, Feng DL.
\newblock {Spatially modulated 'Mottness' in La$_{2-x}$Ba$_x$CuO$_4$}.
\newblock {\em Nat. Phys.\/} {\bf 1} (2005) 155--158.
\newblock \doi{10.1038/nphys178}.

\bibitem[{Zhang and Rice(1988)}]{Zhang1988}
Zhang FC, Rice TM.
\newblock Effective hamiltonian for the superconducting cu oxides.
\newblock {\em Phys. Rev. B\/} {\bf 37} (1988) 3759--3761.
\newblock \doi{10.1103/PhysRevB.37.3759}.

\bibitem[{Green et~al.(2016)Green, Haverkort, and Sawatzky}]{Green2016}
Green RJ, Haverkort MW, Sawatzky GA.
\newblock Bond disproportionation and dynamical charge fluctuations in the
  perovskite rare-earth nickelates.
\newblock {\em Phys. Rev. B\/} {\bf 94} (2016) 195127.
\newblock \doi{10.1103/PhysRevB.94.195127}.

\bibitem[{Been et~al.(2021)Been, Lee, Hwang, Cui, Zaanen, Devereaux
  et~al.}]{Been2021}
Been E, Lee WS, Hwang HY, Cui Y, Zaanen J, Devereaux T, et~al.
\newblock Electronic structure trends across the rare-earth series in
  superconducting infinite-layer nickelates.
\newblock {\em Phys. Rev. X\/} {\bf 11} (2021) 011050.
\newblock \doi{10.1103/physrevx.11.011050}.

\bibitem[{Nomura et~al.(2019)Nomura, Hirayama, Tadano, Yoshimoto, Nakamura, and
  Arita}]{Nomura2019}
Nomura Y, Hirayama M, Tadano T, Yoshimoto Y, Nakamura K, Arita R.
\newblock Formation of a two-dimensional single-component correlated electron
  system and band engineering in the nickelate superconductor
  ${\mathrm{ndnio}}_{2}$.
\newblock {\em Phys. Rev. B\/} {\bf 100} (2019) 205138.
\newblock \doi{10.1103/PhysRevB.100.205138}.

\bibitem[{Botana and Norman(2020)}]{Botana2020}
Botana A, Norman M.
\newblock Similarities and differences between {LaNiO$_2$} and {CaCuO$_2$} and
  implications for superconductivity.
\newblock {\em Phys. Rev. X\/} {\bf 10} (2020) 011024.
\newblock \doi{10.1103/physrevx.10.011024}.

\bibitem[{Kitatani et~al.(2020)Kitatani, Si, Janson, Arita, Zhong, and
  Held}]{Kitatani2020}
Kitatani M, Si L, Janson O, Arita R, Zhong Z, Held K.
\newblock {Nickelate superconductors—a renaissance of the one-band Hubbard
  model}.
\newblock {\em npj Quantum Mater.\/} {\bf 5} (2020) 59.
\newblock \doi{10.1038/s41535-020-00260-y}.

\bibitem[{Wu et~al.(2020)Wu, Di~Sante, Schwemmer, Hanke, Hwang, Raghu
  et~al.}]{Wu2020}
Wu X, Di~Sante D, Schwemmer T, Hanke W, Hwang HY, Raghu S, et~al.
\newblock Robust ${d}_{{x}^{2}\ensuremath{-}{y}^{2}}$-wave superconductivity of
  infinite-layer nickelates.
\newblock {\em Phys. Rev. B\/} {\bf 101} (2020) 060504.
\newblock \doi{10.1103/PhysRevB.101.060504}.

\bibitem[{Zhang et~al.(2020{\natexlab{a}})Zhang, Jin, Wang, Xi, Shi, Ye
  et~al.}]{Zhang2020E}
Zhang H, Jin L, Wang S, Xi B, Shi X, Ye F, et~al.
\newblock Effective hamiltonian for nickelate oxides
  {${\mathrm{Nd}}_{1\ensuremath{-}x}{\mathrm{Sr}}_{x}{\mathrm{NiO}}_{2}$}.
\newblock {\em Phys. Rev. Research\/} {\bf 2} (2020{\natexlab{a}}) 013214.
\newblock \doi{10.1103/PhysRevResearch.2.013214}.

\bibitem[{Sakakibara et~al.(2020)Sakakibara, Usui, Suzuki, Kotani, Aoki, and
  Kuroki}]{Sakakibara2020}
Sakakibara H, Usui H, Suzuki K, Kotani T, Aoki H, Kuroki K.
\newblock Model construction and a possibility of cupratelike pairing in a new
  ${d}^{9}$ nickelate superconductor
  {$(\mathrm{Nd},\mathrm{Sr}){\mathrm{NiO}}_{2}$}.
\newblock {\em Phys. Rev. Lett.\/} {\bf 125} (2020) 077003.
\newblock \doi{10.1103/PhysRevLett.125.077003}.

\bibitem[{Lechermann(2020)}]{Lechermann2020}
Lechermann F.
\newblock Late transition metal oxides with infinite-layer structure:
  Nickelates versus cuprates.
\newblock {\em Phys. Rev. B\/} {\bf 101} (2020) 081110.
\newblock \doi{10.1103/PhysRevB.101.081110}.

\bibitem[{Wang et~al.(2020)Wang, Kang, Miao, and Kotliar}]{Wang2020E}
Wang Y, Kang CJ, Miao H, Kotliar G.
\newblock Hund's metal physics: From {${\mathrm{SrNiO}}_{2}$ to
  ${\mathrm{LaNiO}}_{2}$}.
\newblock {\em Phys. Rev. B\/} {\bf 102} (2020) 161118.
\newblock \doi{10.1103/PhysRevB.102.161118}.

\bibitem[{Werner and Hoshino(2020)}]{Werner2020}
Werner P, Hoshino S.
\newblock Nickelate superconductors: Multiorbital nature and spin freezing.
\newblock {\em Phys. Rev. B\/} {\bf 101} (2020) 041104.
\newblock \doi{10.1103/PhysRevB.101.041104}.

\bibitem[{Adhikary et~al.(2020)Adhikary, Bandyopadhyay, Das, Dasgupta, and
  Saha-Dasgupta}]{Adhikary2020}
Adhikary P, Bandyopadhyay S, Das T, Dasgupta I, Saha-Dasgupta T.
\newblock Orbital-selective superconductivity in a two-band model of
  infinite-layer nickelates.
\newblock {\em Phys. Rev. B\/} {\bf 102} (2020) 100501.
\newblock \doi{10.1103/PhysRevB.102.100501}.

\bibitem[{Karp et~al.(2020)Karp, Botana, Norman, Park, Zingl, and
  Millis}]{Karp2020}
Karp J, Botana AS, Norman MR, Park H, Zingl M, Millis A.
\newblock Many-body electronic structure of {${\mathrm{NdNiO}}_{2}$ and
  ${\mathrm{CaCuO}}_{2}$}.
\newblock {\em Phys. Rev. X\/} {\bf 10} (2020) 021061.
\newblock \doi{10.1103/PhysRevX.10.021061}.

\bibitem[{Liu et~al.(2021)Liu, Xu, Cao, Zhu, Wang, and Yang}]{Liu2021}
Liu Z, Xu C, Cao C, Zhu W, Wang ZF, Yang J.
\newblock Doping dependence of electronic structure of infinite-layer
  {${\mathrm{NdNiO}}_{2}$}.
\newblock {\em Phys. Rev. B\/} {\bf 103} (2021) 045103.
\newblock \doi{10.1103/PhysRevB.103.045103}.

\bibitem[{Wan et~al.(2021)Wan, Ivanov, Resta, Leonov, and Savrasov}]{Wan2021}
Wan X, Ivanov V, Resta G, Leonov I, Savrasov SY.
\newblock Exchange interactions and sensitivity of the ni two-hole spin state
  to hund's coupling in doped {${\mathrm{NdNiO}}_{2}$}.
\newblock {\em Phys. Rev. B\/} {\bf 103} (2021) 075123.
\newblock \doi{10.1103/PhysRevB.103.075123}.

\bibitem[{Lang et~al.(2021)Lang, Jiang, and Ku}]{Lang2021}
Lang ZJ, Jiang R, Ku W.
\newblock Strongly correlated doped hole carriers in the superconducting
  nickelates: Their location, local many-body state, and low-energy effective
  hamiltonian.
\newblock {\em Phys. Rev. B\/} {\bf 103} (2021) L180502.
\newblock \doi{10.1103/PhysRevB.103.L180502}.

\bibitem[{Higashi et~al.(2021)Higashi, Winder, Kune\ifmmode~\check{s}\else
  \v{s}\fi{}, and Hariki}]{Higashi2021}
Higashi K, Winder M, Kune\ifmmode~\check{s}\else \v{s}\fi{} J, Hariki A.
\newblock Core-level x-ray spectroscopy of infinite-layer nickelate:
  $\mathrm{LDA}+\mathrm{DMFT}$ study.
\newblock {\em Phys. Rev. X\/} {\bf 11} (2021) 041009.
\newblock \doi{10.1103/PhysRevX.11.041009}.

\bibitem[{Phillips(2006)}]{Phillips2006}
Phillips P.
\newblock Mottness.
\newblock {\em Ann. Phys. (N. Y.)\/} {\bf 321} (2006) 1634--1650.
\newblock \doi{https://doi.org/10.1016/j.aop.2006.04.003}.

\bibitem[{Jiang et~al.(2020)Jiang, Berciu, and Sawatzky}]{Jiang2020}
Jiang M, Berciu M, Sawatzky GA.
\newblock Critical nature of the ni spin state in doped
  {${\mathrm{NdNiO}}_{2}$}.
\newblock {\em Phys. Rev. Lett.\/} {\bf 124} (2020) 207004.
\newblock \doi{10.1103/PhysRevLett.124.207004}.

\bibitem[{Zhang et~al.(2020{\natexlab{b}})Zhang, Yang, and Zhang}]{Zhang2020S}
Zhang GM, Yang Yf, Zhang FC.
\newblock Self-doped mott insulator for parent compounds of nickelate
  superconductors.
\newblock {\em Phys. Rev. B\/} {\bf 101} (2020{\natexlab{b}}) 020501.
\newblock \doi{10.1103/PhysRevB.101.020501}.

\bibitem[{Hu and Wu(2019)}]{Hu2019}
Hu LH, Wu C.
\newblock Two-band model for magnetism and superconductivity in nickelates.
\newblock {\em Phys. Rev. Research\/} {\bf 1} (2019) 032046.
\newblock \doi{10.1103/PhysRevResearch.1.032046}.

\bibitem[{Liu et~al.(2020)Liu, Ren, Zhu, Wang, and Yang}]{Liu2020}
Liu Z, Ren Z, Zhu W, Wang Z, Yang J.
\newblock Electronic and magnetic structure of infinite-layer {NdNiO$_2$}:
  trace of antiferromagnetic metal.
\newblock {\em npj Quantum Mater.\/} {\bf 5} (2020) 31.
\newblock \doi{10.1038/s41535-020-0229-1}.

\bibitem[{Katukuri et~al.(2020)Katukuri, Bogdanov, Weser, van~den Brink, and
  Alavi}]{Katukuri2020}
Katukuri VM, Bogdanov NA, Weser O, van~den Brink J, Alavi A.
\newblock {Electronic correlations and magnetic interactions in infinite-layer
  ${\mathrm{NdNiO}}_{2}$}.
\newblock {\em Phys. Rev. B\/} {\bf 102} (2020) 241112.
\newblock \doi{10.1103/PhysRevB.102.241112}.

\bibitem[{Lu et~al.(2018)Lu, Betto, F\"ursich, Suzuki, Kim, Cristiani
  et~al.}]{Lu2018}
Lu Y, Betto D, F\"ursich K, Suzuki H, Kim HH, Cristiani G, et~al.
\newblock Site-selective probe of magnetic excitations in rare-earth nickelates
  using resonant inelastic x-ray scattering.
\newblock {\em Phys. Rev. X\/} {\bf 8} (2018) 031014.
\newblock \doi{10.1103/PhysRevX.8.031014}.

\bibitem[{Hutchings and Samuelsen(1972)}]{Hutching1972}
Hutchings MT, Samuelsen EJ.
\newblock {Measurement of Spin-Wave Dispersion in NiO by Inelastic Neutron
  Scattering and Its Relation to Magnetic Properties}.
\newblock {\em Phys. Rev. B\/} {\bf 6} (1972) 3447--3461.
\newblock \doi{10.1103/PhysRevB.6.3447}.

\bibitem[{Ghiringhelli et~al.(2009)Ghiringhelli, Piazzalunga, Dallera, Schmitt,
  Strocov, Schlappa et~al.}]{Ghiringhelli2009}
Ghiringhelli G, Piazzalunga A, Dallera C, Schmitt T, Strocov VN, Schlappa J,
  et~al.
\newblock {Observation of Two Nondispersive Magnetic Excitations in NiO by
  Resonant Inelastic Soft-X-Ray Scattering}.
\newblock {\em Phys. Rev. Lett.\/} {\bf 102} (2009) 027401.
\newblock \doi{10.1103/PhysRevLett.102.027401}.

\bibitem[{Peng et~al.(2018)Peng, Huang, Fumagalli, Minola, Wang, Sun
  et~al.}]{Peng2018}
Peng YY, Huang EW, Fumagalli R, Minola M, Wang Y, Sun X, et~al.
\newblock {Dispersion, damping, and intensity of spin excitations in the
  monolayer
  ${(\text{Bi,Pb})}_{2}{(\text{Sr,La})}_{2}{\mathrm{CuO}}_{6+\ensuremath{\delta}}$
  cuprate superconductor family}.
\newblock {\em Phys. Rev. B\/} {\bf 98} (2018) 144507.
\newblock \doi{10.1103/PhysRevB.98.144507}.

\bibitem[{Jia et~al.(2014)Jia, Nowadnick, Wohlfeld, Kung, Chen, Johnston
  et~al.}]{Jia2014}
Jia CJ, Nowadnick EA, Wohlfeld K, Kung YF, Chen CC, Johnston S, et~al.
\newblock Persistent spin excitations in doped antiferromagnets revealed by
  resonant inelastic light scattering.
\newblock {\em Nat. Commun.\/} {\bf 5} (2014) 3314.
\newblock \doi{10.1038/ncomms4314}.

\bibitem[{Bala et~al.(1995)Bala, Ole\ifmmode~\acute{s}\else \'{s}\fi{}, and
  Zaanen}]{Bala1995}
Bala J, Ole\ifmmode~\acute{s}\else \'{s}\fi{} AM, Zaanen J.
\newblock Spin polarons in the t-t\ensuremath{'}-j model.
\newblock {\em Phys. Rev. B\/} {\bf 52} (1995) 4597--4606.
\newblock \doi{10.1103/PhysRevB.52.4597}.

\bibitem[{Scalapino(2012)}]{Scalapino2012}
Scalapino DJ.
\newblock A common thread: The pairing interaction for unconventional
  superconductors.
\newblock {\em Rev. Mod. Phys.\/} {\bf 84} (2012) 1383--1417.
\newblock \doi{10.1103/RevModPhys.84.1383}.

\bibitem[{Poltavets et~al.(2010)Poltavets, Lokshin, Nevidomskyy, Croft, Tyson,
  Hadermann et~al.}]{Poltavets2010}
Poltavets VV, Lokshin KA, Nevidomskyy AH, Croft M, Tyson TA, Hadermann J,
  et~al.
\newblock {Bulk Magnetic Order in a Two-Dimensional
  ${\mathrm{Ni}}^{1+}/{\mathrm{Ni}}^{2+}$ (${d}^{9}/{d}^{8}$) Nickelate,
  Isoelectronic with Superconducting Cuprates}.
\newblock {\em Phys. Rev. Lett.\/} {\bf 104} (2010) 206403.
\newblock \doi{10.1103/PhysRevLett.104.206403}.

\bibitem[{Botana et~al.(2016)Botana, Pardo, Pickett, and Norman}]{Botana2016}
Botana AS, Pardo V, Pickett WE, Norman MR.
\newblock {Charge ordering in ${\mathrm{Ni}}^{1+}/{\mathrm{Ni}}^{2+}$
  nickelates: ${\mathrm{La}}_{4}{\mathrm{Ni}}_{3}{\mathrm{O}}_{8}$ and
  ${\mathrm{La}}_{3}{\mathrm{Ni}}_{2}{\mathrm{O}}_{6}$}.
\newblock {\em Phys. Rev. B\/} {\bf 94} (2016) 081105.
\newblock \doi{10.1103/PhysRevB.94.081105}.

\bibitem[{Zhang et~al.(2019)Zhang, Pajerowski, Botana, Zheng, Harriger,
  Rodriguez-Rivera et~al.}]{Zhang2019}
Zhang J, Pajerowski DM, Botana AS, Zheng H, Harriger L, Rodriguez-Rivera J,
  et~al.
\newblock Spin stripe order in a square planar trilayer nickelate.
\newblock {\em Phys. Rev. Lett.\/} {\bf 122} (2019) 247201.
\newblock \doi{10.1103/PhysRevLett.122.247201}.

\bibitem[{Carlson et~al.(2004)Carlson, Yao, and Campbell}]{Carlson2004spin}
Carlson EW, Yao DX, Campbell DK.
\newblock Spin waves in striped phases.
\newblock {\em Phys. Rev. B\/} {\bf 70} (2004) 064505.
\newblock \doi{10.1103/PhysRevB.70.064505}.

\bibitem[{Huangfu et~al.(2020)Huangfu, Guguchia, Cheptiakov, Zhang, Luetkens,
  Gawryluk et~al.}]{Huangfu2020}
Huangfu S, Guguchia Z, Cheptiakov D, Zhang X, Luetkens H, Gawryluk DJ, et~al.
\newblock {Short-range magnetic interactions and spin-glass behavior in the
  quasi-two-dimensional nickelate Pr$_4$Ni$_3$O$_8$}.
\newblock {\em Phys. Rev. B\/} {\bf 102} (2020) 054423.
\newblock \doi{10.1103/physrevb.102.054423}.

\bibitem[{Ortiz et~al.(2021)Ortiz, Menke, Misj\'ak, Mantadakis, F\"ursich,
  Schierle et~al.}]{Ortiz2021}
Ortiz RA, Menke H, Misj\'ak F, Mantadakis DT, F\"ursich K, Schierle E, et~al.
\newblock Superlattice approach to doping infinite-layer nickelates.
\newblock {\em Phys. Rev. B\/} {\bf 104} (2021) 165137.
\newblock \doi{10.1103/PhysRevB.104.165137}.

\end{thebibliography}

\end{document}